# Using Site Testing Data for Adaptive Optics Simulations

Glen Herriot[a], David Andersen[a], Rod Conan[d], Brent Ellerbroek[b], Luc Gilles[b], Paul Hickson[c], Kate Jackson[d], Olivier Lardière[d], Thomas Pfrommer[c], Jean-Pierre Véran[a], Lianqi Wang[b]

[a]Herzberg Institute of Astrophysics, 5071 W. Saanich Rd., Victoria, Canada V9E 2E7;
[b]TMT Observatory Corp., 2632 E. Washington Blvd., Pasadena, CA 91107;
[c]Dept. of Physics & Astronomy, University of British Columbia, 6224 Agricultural Rd., Vancouver, Canada, V6T 1Z1
[d]Adaptive Optics Lab, ELW A212, University of Victoria, Victoria, Canada, V8W 3P6.

## ABSTRACT

Astronomical Site testing data plays a vital role in the simulation, design, evaluation and operation of adaptive optics systems for large telescope. We present the example of TMT and its first light facilitiy adaptive optics system NFIRAOS, and illustrate the many simulations done based on site testing data.

## 1. INTRODUCTION

Site survey data serves three important roles in guiding the design and operation of an adaptive optics system:

1. To help dimension the system
2. To predict performance on-sky
3. To provide constraints and prior information to assist the real time control system during observations

Several of the early-light suite of instruments for the Thirty Meter Telescope (TMT) [1][2] will be fed by a facility adaptive optics system, NFIRAOS[3] (Narrow Field Infrared Adaptive Optics System) that compensates atmospheric turbulence. The performance requirements for NFIRAOS are given in Section 2. The initial version of NFIRAOS will have diffraction-limited performance in the near IR with 50 per cent sky coverage at the galactic pole. This paper begins with a brief description of NFIRAOS, and gives examples of how we use site survey data to guide the design of NFIRAOS.

The left half of Figure 1 shows the space allowed for NFIRAOS on one of the two Nasmyth platforms of the TMT, with the telescope primary mirror in the background. The input beam from the telescope arrives from the tertiary mirror, visible in the center of the main 30-m mirror. NFIRAOS corrects the science light and supports up to three client instruments shown here as cylinders on the top, bottom and the side in the right half of Figure 1. The dark grey exoskeleton carries a total of 47 T for NFIRAOS, three instruments and a science calibration light source [7] (light green.) The cooled electronics enclosure also sits on the Nasmyth platform.

It first outlines key site testing parameters and then continues with examples of the resulting benefit for Adaptive Optics Simulations including: sky coverage; performance models versus season and site; DM Stroke requirements; diameter of Laser launch telescope. Another valuable aspect of measurements is the impact of sodium layer structure and its variability. This paper discusses simulation of making centroiding algorithms resistant to these variations, as well as to assess the effect of sporadic meteor trails on laser guide star adaptive optics. Finally we summarize an autoregressive model of seeing and how it is used to simulate background optimization tasks in NFIRAOS' real time computer, in particular centroid gain estimation for the Natural-guide-star mode wavefront sensor.

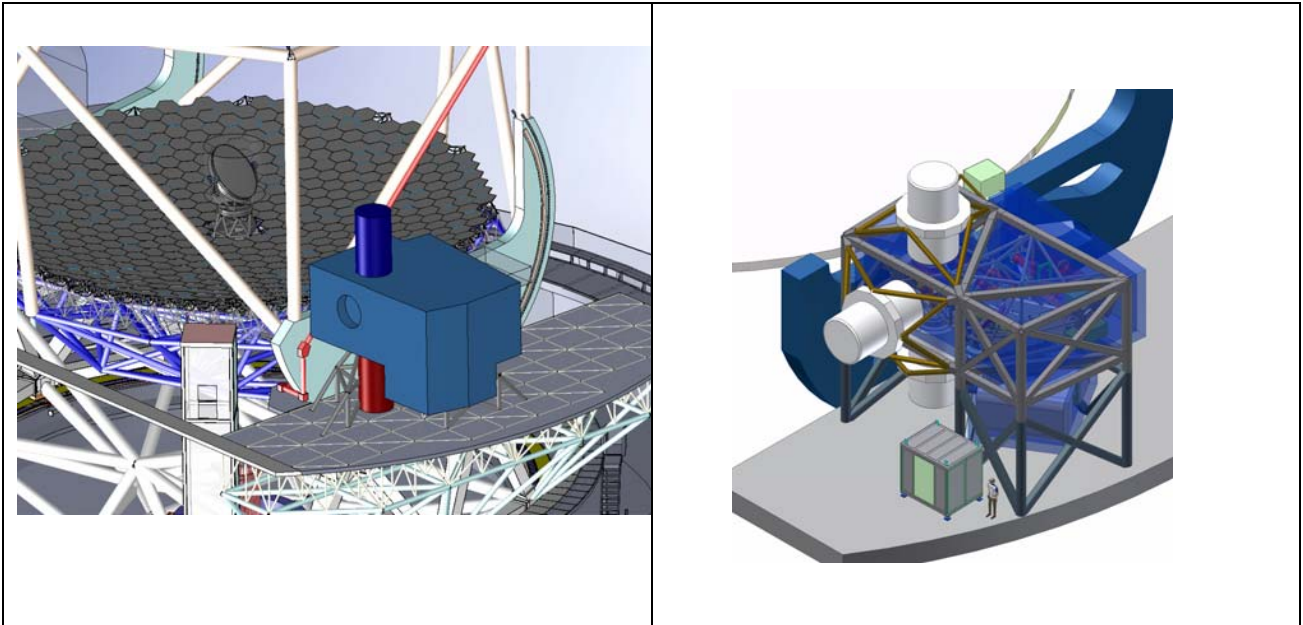

Figure 1 NFIRAOS on Nasmyth Platform of TMT

## 2. TOP-LEVEL NFIRAOS DESIGN REQUIREMENTS

The top-level requirements documents for the TMT observatory specify that the facility AO system shall have:

- 85 per cent throughput from 0.8 to 2.5 $\mu$m
- Thermal emission < 15 per cent of background from sky and telescope
- 187 nm RMS wavefront error on-axis, and tilt-removed WFE of 190 nm on a 30" field of view (FoV)
- High enclosed energy within 160 mas pixels over a 2' FoV
- 50 per cent sky coverage at the Galactic pole
- 2% differential photometry for a 10 minute exposure on a 30" FoV
- 50 $\mu$as differential astrometry for a 100s exposure on a 30" FoV
    - Error falling as $t^{-1/2}$ to a systematic floor of 10 mas
- System available from standby within 10 minutes
- 5 minutes to acquire a new field
- < 1 per cent unscheduled downtime

## 3. NFIRAOS ARCHITECTURE

To meet the above requirements, NFIRAOS has the following features. It is an order 60x60 multi-conjugate AO system with two deformable mirrors optically conjugate to ranges of 0 and 11.2 km, six high-order wavefront sensors observing laser guide stars in the mesospheric sodium layer, and several low-order, infrared natural guide star wavefront sensors located within each client instrument. Among these on-instrument sensors, there will be at least one tip/tilt/focus sensor, and most instruments will also have an additional pair of tip/tilt sensors.

In order to achieve high image quality uniformly over the field of view in the near infrared, NIFROAS employs tomographic wavefront reconstruction.

The 50% sky coverage requirement, at the galactic pole is in conjunction with simultaneously meeting the specification on wavefront error. NFIRAOS will use near-infrared tip/tilt and focus sensing. The tip/tilt sensors are within instruments and use infrared stars sharpened by the adaptive optics system, so their diffraction limited image cores can be sensed accurately with better signal to noise. As well, red stars are more numerous and less obscured by dust. We expect that the majority of the time, NFIRAOS will be guiding using M stars. The wide 2 arcminute guide field of view corrected by dual deformable mirrors assists achieving high sky coverage by sharpening guide stars over the full guide field.

Good optical throughput and low background result from a science path design with a minimum number of surfaces, cooled to minimize emission. Recently to reduce image distortion in the science path to 0.0017%, we have revised the optical design. We have in series, two relays, each with two off-axis paraboloidal mirrors. One DM is located in each relay. As shown in Figure 2 the light from TMT enters via an evacuated dual-pane window and is nearly collimated by off-axis paraboloid 1. The first deformable mirror DM11 is conjugate to 11.2 km in front of the telescope. The second, DM0, resides on a tip/tilt stage. After DM11, the science light passes through a beamsplitter before being re-imaged by OAP4. Finally an instrument selector fold mirror directs the light to instruments on the top, bottom or side ports.

**Figure 2** shows all of the cold NFIRAOS optics from the same viewpoint as the previous figure. The science path is in red. The science beamsplitter diverts visible light downwards from the science path where it is immediately split into a natural visible light path shown in green, and the artificial laser light shown in yellow. This laser light is reimaged by an OAP, second from the left in the figure, and then sent into the laser wavefront sensor trombone optics.

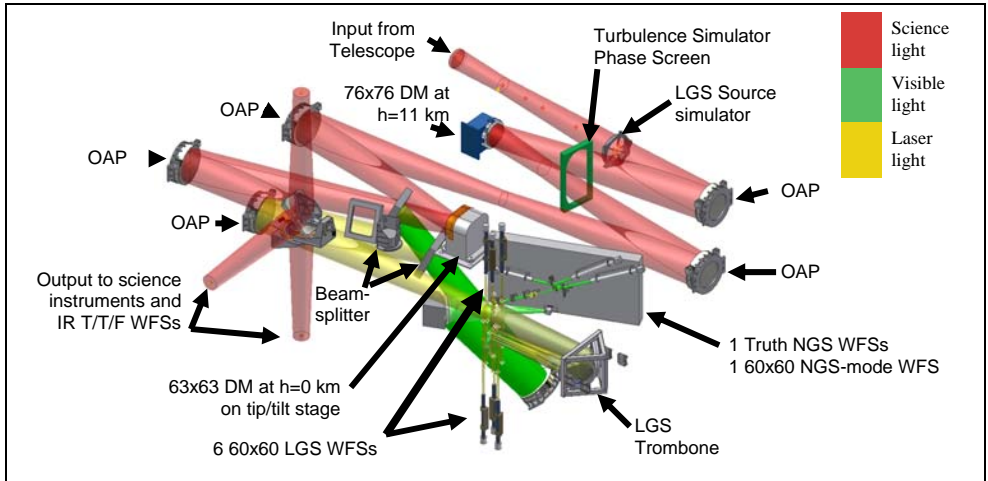

**Figure 2  NFIRAOS Optical paths**

The natural visible light in green continues downwards through the second beamsplitter and is reflected by a copy of OAP4 that reimages the beam. The reimaged natural visible light is sent to the vertical optical bench containing a collection of visible NGS wave-front sensors described later.

As noted above, there are three copies of OAP4 in NFIRAOS: for reimaging the science, laser and natural visible light. This approach permits placing beamsplitters in collimated light to reduce aberrations, in particular astigmatism. A block diagram of the optics is shown next.

## 4. ATMOSPHERIC PARAMETERS OF INTEREST FOR ADAPTIVE OPTICS

The following parameters from site measurements are useful for adaptive optics simulations.

- $r_0$ seeing and evolution of seeing vs. time
- $\theta_0 ... \theta_n$   Isoplanatic Angle, generalized for $n$ Deformable Mirrors
- $L_0$ Outer scale of turbulence
- $\tau_0$ time constant for turbulence evolution
- $C_n^2$ versus altitude

- Time evolution of atmospheric Layers' strength of turbulence
- Wind speed versus altitude
- Ground Level Wind-speed to assess telescope windshake versus dome seeing
- Sodium layer structure, abundance and time variation
- Ground level temperature and variation with time
- Sky transparency versus time

### 4.1 Seeing $r_0$

Seeing $r_0$ is the critical parameter for dimensioning an adaptive optics system. Seeing affects the choice of: the number of actuators needed on DMs; stroke needed on DM on actuators; the number of subapertures on wavefront sensors; laser guide star power required; sky coverage (probability of achieving specified image quality); and computing power in real time computer.

The time evolution of $r_0$ affects the update rate and accuracy of background tasks to optimize adaptive optics control loops.

### 4.2 Outer scale of turbulence $L_0$

The outer scale of turbulence influences the amount of DM stroke required .Smaller L0 means less energy in low-order modes and low temporal frequencies. Smaller $L_0$ means less stroke is needed for the same $r_0$. Similarly $L_0$ affects the stroke and bandwidth of Tip/Tilt mirrors. As well $L_0$ influence the scale of turbulence phase screens used for both optical numerical simulations. The time evolution of $L_0$ affects background tasks, which optimize Adaptive optics control loops.

### 4.3 Isoplanatic Angle $\theta_0$

The isoplanatic angle, $\theta_0$ can be generalized to $\theta_N$ for an adaptive optics system with N deformable mirrors. In this generalized case it refers to the width of the corrected field of view, rather than the natural seeing isoplanatic angle. Thus, $\theta_N$ affects sky coverage because tip/tilt/focus stars should be found within the corrected field in order to take advantage of smaller images to better measure tip/tilt. Furthermore, it influences the optimal number of DMs, and their ideal altitude of conjugation. $\theta_N$ is also a factor in determining the number of Laser Guide Stars and their spacing on the sky.

Finally $\theta_N$ determines the number and location of optical phase screens in a turbulence simulator.

### 4.4 $\tau_0$ time constant of turbulence evolution

$\tau_0$ affects the bandwidth for an AO control system, including the readout rate and thus read noise of wavefront sensors; laser power; and computer speed of real time controller.

### 4.5 $C_n^2$ vs altitude

$C_n^2$ determines the number of layers in tomographic reconstruction within a real time computer, and thus the computing power needed. It defines DM quantity and the optimal altitude of conjugation. Good $C_n^2$ data may be used in simulations to determine the potential effectiveness of a Ground Layer AO system. Equally importantly, good initial $C_n^2$ data, when turning on an AO system, allows quick convergence of tomography algorithm before beginning a science exposure.

### 4.6 Wind speed versus altitude

Under the assumption of frozen flow, predictive filter methods are desirable in a real time controller. In principle, these techniques, while computationally intensive, can reduce servo lag errors in AO systems. But how effective are they? Simulations can tell us, providing that we have good data.

Ground-level wind speed data is used to feed Computational Fluid Dynamics (CFD) dome models to assess wind forces [12], which are applied to TMT structural finite element models and controls model of telescope and mirror segments. The resulting windshake is a disturbance input to NFIRAOS simulations of performance and sky coverage. We ran simulations of adaptive vibration cancellation algorithms with inputs from windshake, machinery vibration and atmospheric turbulence. The advanced controller can reduce 23 milli-arcsecond (mas) RMS tip/tilt jitter to below 0.1 mas.

Dome and mirror flushing (driven by ground level winds) is simulated by computational fluid dynamics and heat transfer models to create dome seeing voxel (volume elements) maps within the dome. Ray tracing through dome voxels creates phase screens, which are also input to Adaptive Optics simulations.

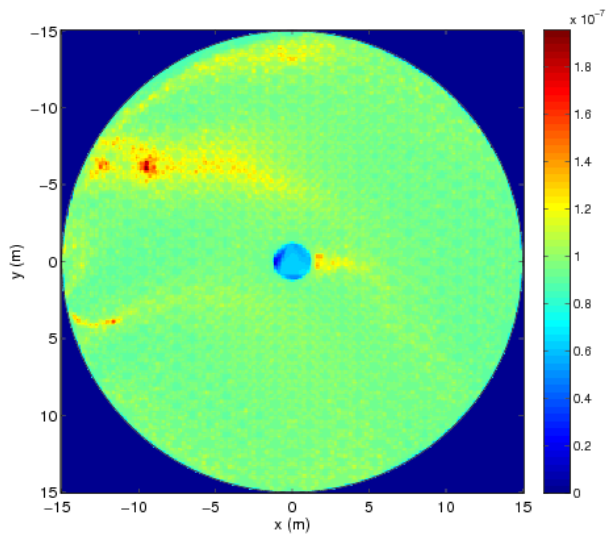 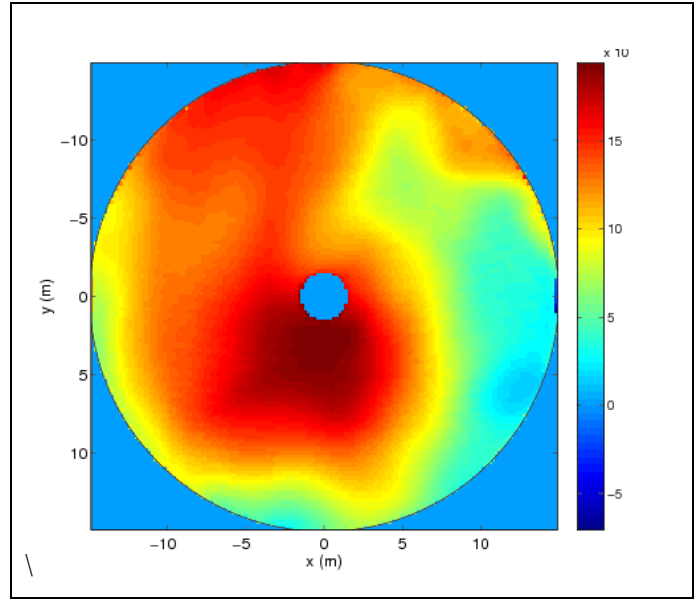

Figure 3 Mirror Seeing (left); Dome seeing Right

Figure 3 shows an extreme case of mirror seeing, showing the layer of voxel elements close to the primary mirror, for an extreme case of having the mirror 20 C away from ambient temperature. The residual wavefront error from this disturbance, after correction by NFIRAOS is 14 nm, out of the total NFIRAOS error budget of 187 nm rms. The right half of Figure 3 shows dome seeing Optical Path Differences (OPDs) sampled at 1/5 m, generated every 0.1 seconds. Simulations were conducted for 5 different time frames, and resulted in a residual wavefront error of 16 nm rms.

### 4.7 Ground level temperature variation with time

Temperature variations of the telescope and dome cause dome seeing as shown above. As well, Near-IR background flux from warm telescope optics increases the integration time for background limited objects, thus affecting the point source sensitivity calculations for TMT. TMT has developed detailed simulations representing an average year of observing that combines the orientation of the telescope in Altitude and Azimuth, based on Gemini observing logs, with time series of wind speed, wind direction and temperature. This simulation determines the temperature of surfaces and voxels within the dome and in the disturbed airflow over the dome.[11]

### 4.8 Fraction of nights with Sub-visible cirrus.

Sub-visible cirrus clouds cause fratricide among laser beacons, and also reduce the returned laser flux due to scattering. Four scattering effects were studied: Rayleigh, ozone, aerosol, and cirrus.[9]

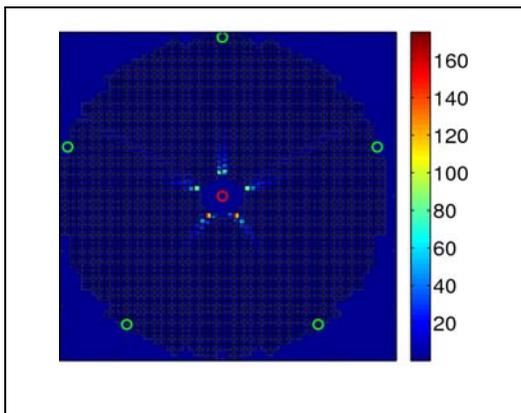

Figure 4 Scattering in central LGS WFS

Rayleigh scattering induces fratricide between LGS WFSs, especially if the laser beacons are centrally launched from behind the secondary mirror. Figure 4 shows the background intensity seen in each subaperture of the central LGS WFS for NFIRAOS. The five radial stripes are due to scattering from the upwards transmission of lasers creating the other five laser beacons. This image was computed by integrating the atmospheric backscattered light intensity profile (volume scattering coefficient) over altitude, subaperture area, and pixel field of view. Real-time updates at ~0.1Hz are expected to provide required bias calibration accuracy to better than 80%

As well, ozone, aerosol and cirrus contribute to momentary signal level variations causing ~23 nm RMS errors for 20% reduction in returned flux.

# 5. TELEMETRY FROM AO "SURVEYS" SITE.

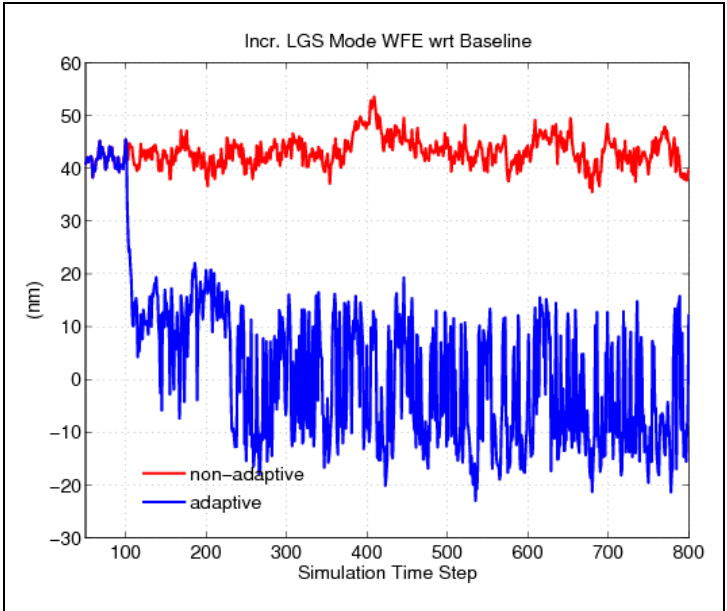

**Figure 5** Tomographic errors with real-time $C_n^2$ estimation

Telemetry from Adaptive Optics Systems can continue to monitor sites. For example Altair, the classic (single DM and single WFS) AO System on Gemini North outputs $r_0$ and $L_0$ based on telemetry. For the Gemini Gpi AO system under construction, Poyneer & Véran ran simulations using Gemini Altair and Gemini NICI telemetry. [13]These simulations indicate that GPi can determine the number of atmospheric layers and wind speed for each. But it cannot determine the altitude and strength of each layer.

While there is a good fraction of turbulence that appears to be frozen flow, there is also a significant portion that is not. All currently proposed AO predictive control schemes currently assume frozen flow.

## 5.1 Real-Time $C_n^2$ Profile Estimation

For optimal Tomographic Wavefront Reconstruction, it is helpful to know the $C_n^2$ profile. This prior knowledge can constrain the tomographic solution. Luc Gilles at the TMT project office developed a SLODAR-like method that correlates pseudo open-loop measurements from a pair of the 6 NFIRAOS LGS WFSs. The method eliminates sensitivity to LGS tip/tilt/focus by using second-order differences of gradients. It is computationally efficient and convergent in a few hundred frames at 800 Hz. Six layers are estimated from 11 baselines, giving a vertical resolution of ~1.5 km. Running as a background task that then adaptively updates the tomography, it reduces tomographic errors as shown in Figure 5.

# 6. TMT ERROR BUDGETING AND PERFORMANCE ANALYSIS

TMT is undertaking a comprehensive evaluation of TMT AO architecture.[5] This performance evaluation includes: wavefront disturbances due to the atmosphere, telescope, NFIRAOS, and instruments; NFIRAOS wavefront sensing and correcting hardware; Laser guide star and On-instrument WFS (OIWFS)[10] components; and NFIRAOS processing algorithms.

Performance evaluation is computed as a function of seeing, zenith angle, field of view and galactic latitude. These estimates developed through a combination of:

- Integrated AO simulations
- Side analyses
- Budget allocations
- Lab and LIDAR experiments

Key results for Mauna Kea confirm that performance requirements are met: 187 nm wavefront error on-axis at zenith with median seeing and 50% sky coverage at the Galactic Pole is met with 83 nm RMS margin in quadrature. This conclusion is based upon detailed time domain simulations of NFIRAOS, including WFSs, DMs, RTC, and telescope models.

Sky coverage has also been evaluated and optimized in detail using physical optics modeling of the OIWFSs. As mentioned earlier, sky coverage is the probability of achieving the specified image quality. Fundamentally sky coverage is limited by the availability of natural guide stars that are needed to sense Tip Tilt and Focus with the OIWFSs. OIWFS Pixel processing and temporal filtering algorithms have been studied in detail.

The simulations involve Monte Carlo runs over 500 guide star fields. Performance is evaluated as a function of zenith angle and seeing at 25%, 50% and 75% percentile conditions derived from the TMT site survey data for Mauna Kea. Table 1 shows the median and 25$^{th}$ percentile seeing and Cn2 profiles together with wind speed versus altitude.

| Altitude (km) | 0 | 0.5 | 1 | 2 | 4 | 8 | 16 |
|---|---|---|---|---|---|---|---|
| Wind Speed (m/s) | 5.6 | 5.8 | 6.2 | 7.6 | 13 | 19 | 12 |
| **MK13N 25% profile, $r_0$= 27.4 cm, $\theta_0$ =2.7", $f_G$=15.9 Hz** | | | | | | | |
| Weights (%) | 32 | 15 | 4.7 | 4.1 | 16 | 11 | 18 |
| **MK13N 50% profile, $r_0$= 19.9 cm, $\theta_0$ =2.2", $f_G$=21.7 Hz** | | | | | | | |
| Weights (%) | 29 | 18 | 6.6 | 7.8 | 14 | 12 | 13 |

**Table 1** 25$^{th}$ and 50$^{th}$ percentile winds aloft, $r_0$ and $C_n^2$ profiles for MK

Figure 6 shows the point spread function of the expected images delivered by NFIRAOS on TMT for median conditions on Mauna Kea. The left hand panel is an image displayed with a logarithmic intensity scale. The upper right panel is a cut through the central region of the PSFs for J,H, and K bands. The lower right graph is a cut through the same PSFs, but over a wider field of view.

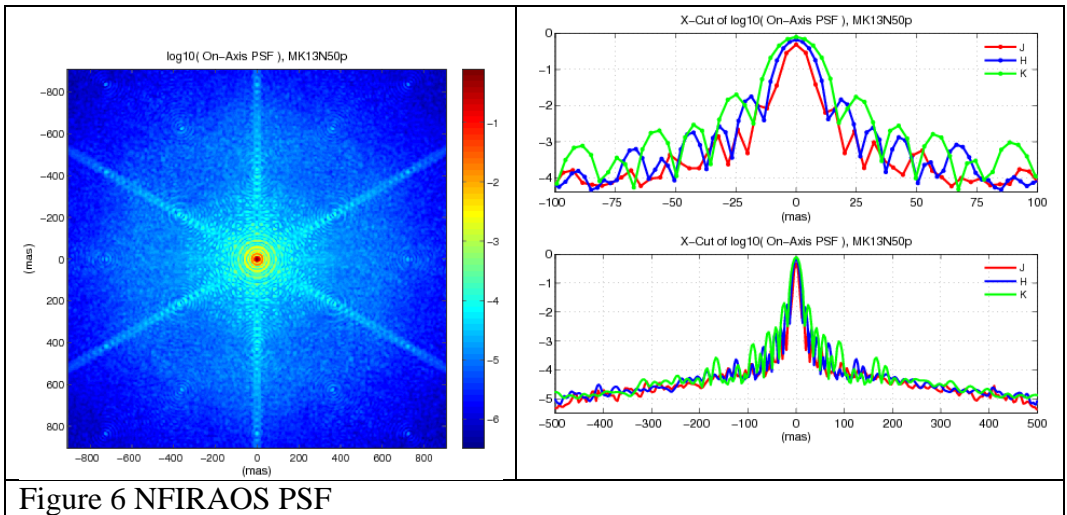

Figure 6 NFIRAOS PSF

# 7. ENCIRCLED ENERGY FOR A MULTISLIT SPECTROGRAPH

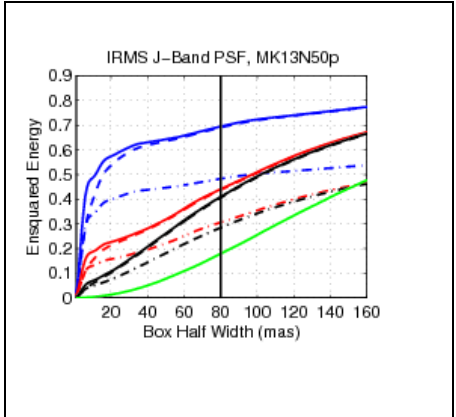

Figure 7 Ensquared Energy

However much of the science planned for TMT, is spectroscopy, not just imaging.[8] Here the figure of merit is the fraction of the energy that falls in the slit of a multi-object spectrograph. In Figure 7, we see a family of curves for H-band under median conditions. The blue curves are for a slit in the centre of the field of view, the red curves are an average over a 1 arcminute field, and the black curves are for a 2' FoV. The green curve is for uncorrected natural seeing under median conditions. The dashed lines are the most realistic expected ensquared energy curves. The solid curve is based on an idealized AO system model that neglects implementation errors such as from NFIRAOS' optics polishing and alignment. The dash-dot line shows the pessimistic result from using the Marechal approximation to add in the implementation errors to the idealized AO system. The dashed, realistic lines use physical optics modeling of all errors, including correcting the atmosphere and the telescope and NFIRAOS.

# 8. SKY COVERAGE ANALYSIS

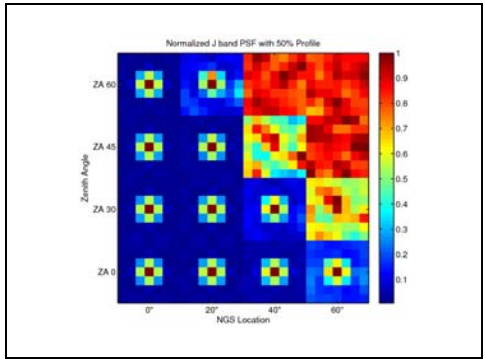

**Figure 8 PSF vs zenith and field angle**

One of the most important considerations for an AO system is the percentage of the sky over which it meets its error budget. In the case of NFIRAOS, the requirement is 50% probability at the galactic pole, where natural guide stars are least numerous. Although NFIRAOS uses laser guide stars to correct high order aberrations, it relies on measurements from three natural stars by the OIWFSs to correct tip, tilt, focus and image distortion errors like plate scale. Away from the pole, sky coverage will improve except in highly obscured or very crowded regions. NFIRAOS relies on the image sharpening from the high-order laser guide star system to produce near-infrared diffraction limited images of natural stars. The sharpened images can be centroided more accurately, using fewer detector pixels, resulting in high signal to noise measurements.

NFIRAOS performance meets its requirements at zenith with a comfortable margin. Unfortunately, off-zenith, strehl ratio, $\theta_0$ and $\theta_2$ all get smaller. We see in the lower left of Figure 8, the PSF for a natural star on-axis when observed at zenith; under the combination of wide field angles and/or high zenith angles, the images break up into speckles.

Since, these images no longer have diffraction limited cores, accurate centroiding is not possible, and one would think that sky coverage suffers. Fortunately, for TMT on Mauna Kea, regions of the sky that have few guide stars reach high elevation angles, and regions that never get high in the sky are rich in guide stars. Hence it is usually possible to choose guide stars at small field angles or observe at high elevation above the horizon.

As well as atmospheric tip/tilt, a large fraction of the image wander is due to windshake of the telescope. Wind measurements from site survey data are processed through CFD (computational fluid dynamics) models of the dome and telescope to produce time series of forces on the telescope. These are then processed by simulations incorporating telescope controls, dynamics and optics to provide image jitter inputs to AO simulations.

## 9. PERFORMANCE VERSUS SEASONS

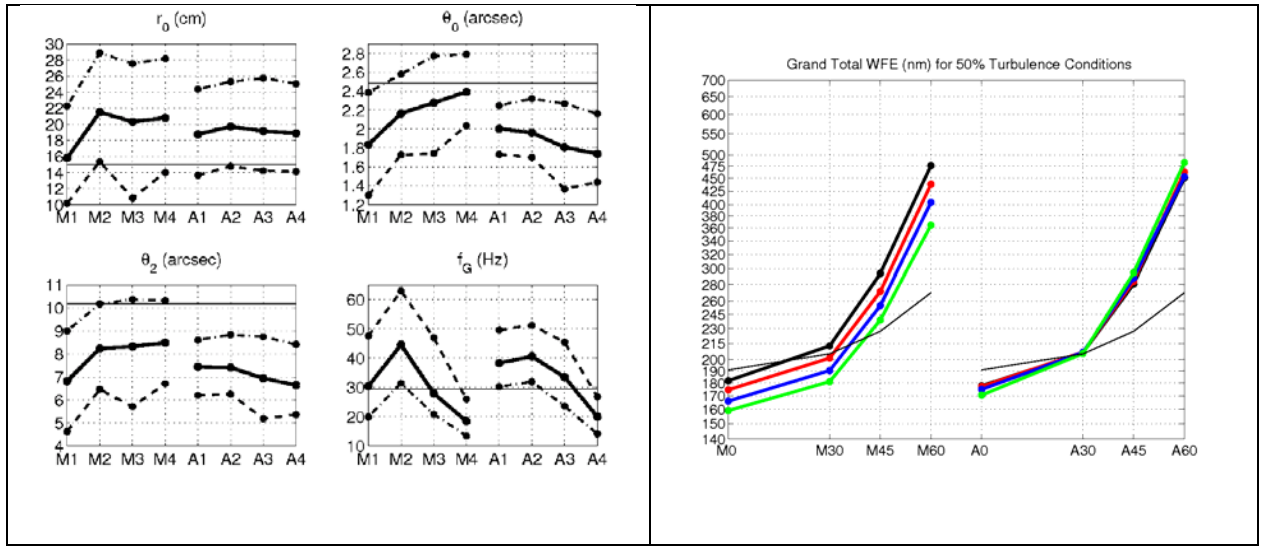

**Figure 9 (Left) Site data; (Right) performance. vs seasons**

The site survey data versus the seasons is shown in the left of Figure 9, showing seeing, isoplanatic angle, generalized isoplanatic angle for 2 DMs, and Greenwood frequency – all at zenith at 500 nm wavelength. The solid curves are for median conditions, the dashed lines show the 25$^{th}$ best and worst percentiles. M1..M4 are the four seasons on Mauna Kea and A1..A4 are the four seasons on Cerro Armazonas. The right hand figure shows wavefront error (WFE) – smaller is better. Black, red, blue, green curves correspond respectively to the winter/spring/summer/fall seasons. The horizontal axis is for Zenith angle for Mauna Kea and Armazonas. The thin black line show expected performance using a standardized $C_n^2$ profile guessed before site survey data became available from the TMT monitoring campaign. In general $r_0$ is better, and $\theta_2$ worse than guessed.

## 10. DM STROKE REQUIREMENTS

Figure 10 shows a histogram of the DM actuator commands, with the Deformable Mirror that is optically conjugate to the ground displayed on the left, and the DM conjugate to 11.2 km shown on the right.

Actuator optical path differences (OPDs or 2x stroke) are shown for the ground and upper DMs for a variety of measured turbulence profiles that have similar 90th percentile $\theta_0$. But these profiles have quite different values of $r_0$, ranging from 0.07 m to 0.193 m. The outer scale is assumed to be 30 m for these simulations.

The upper DM has more or less similar command distributions for all of the profiles, while the ground-conjugate DM has broader histograms for smaller values of $r_0$.

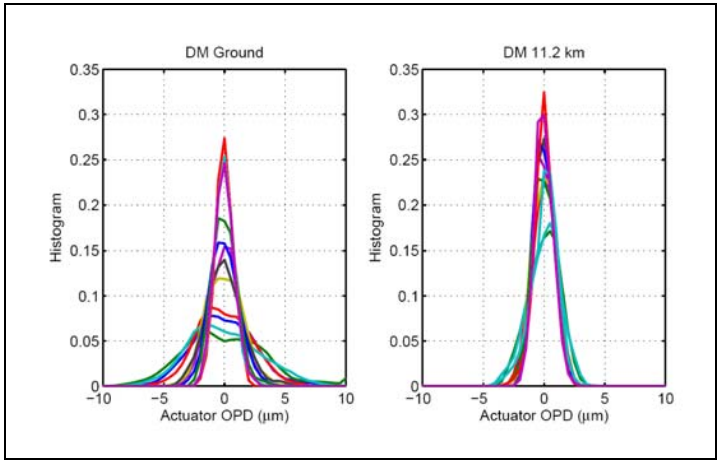
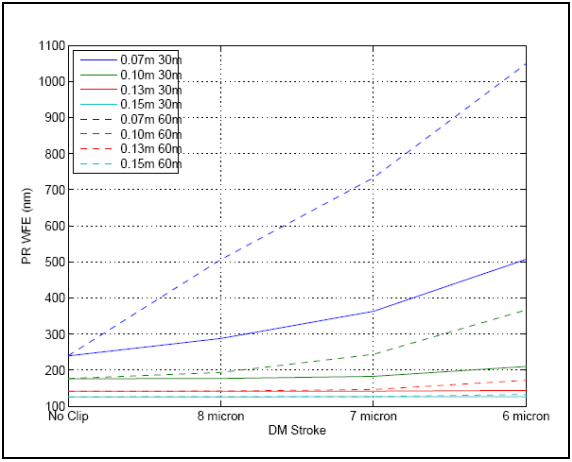

Figure 10 Histogram of DM actuator commands.

Figure 11 DM actuator stroke vs $L_0$ & $r_0$.

Meanwhile, if $L_0$ is large for a given $r_0$, then the DM requires more stroke to achieve the same wavefront same error. The next figure shows an example simulation for a classic adaptive optics system with a single ground-conjugated DM and single WFS. The dashed lines are for $L_0 = 30$ m and the solid curves are for $L_0 = 60$ m. The value of $r_0$ ranges from 0.07 m to 0.15 m.

In each pair of lines for the same $r_0$ (of the same colour) the best performance (lower WFE) occurs when $L_0$ is smallest, unless the DM stroke is increased, which is expensive. The X axis of the graph runs backwards and corresponds to the maximum stroke available from a DM. The left of the axis assumes infinite stroke (no command clipping) and the right end is for a DM that has the 6 microns of stroke.

This graph illustrates why Adaptive Optics designers would like to get accurate measurements of outer scale to avoid spending money needlessly on actuator stroke, while assuring that the AO system can operate in poor seeing.

## 11. SITE SURVEY TEMPERATURE DATA

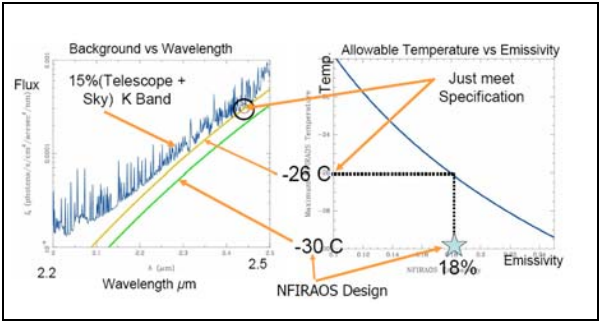

**Figure 12 Temperature vs emissivity NFIRAOS**

Mountain-top temperature drives AO system temperature for low background observations. The requirement for NFIRAOS to add < 15% to the sky and telescope background in K band implies cooling NFIRAOS. Observing time decreases directly with decrease in thermal background. Cooling NFIRAOS cuts observing time by a factor of 2.4 in K band. Since the median Temperature on Mauna Kea is 2.3 C, NFIRAOS operates at -30 C to meet its background requirement.

In Figure 12 on the left, the jagged curve is the specification plotted in K band versus wavelength. The jagged peaks are measured sky OH⁻ lines added to a grey body curve representing the thermal emission from three mirrors of the TMT telescope at median temperature on Mauna Kea, and then scaled by a factor of 15%. If the summit were warmer, then this curve would move up, and NFIRAOS would not have to be so cold to stay under it. The upper smooth curve on the left in yellow shows the emission from NFIRAOS if it operated at – 26 C and just met the specification. The lower green curve shows the performance at the baseline temperature of – 30 C. The right hand figure shows the operating temperature on the vertical axis versus optics emissivity on the horizontal axis.

## 12. TURBULENCE SIMULATOR

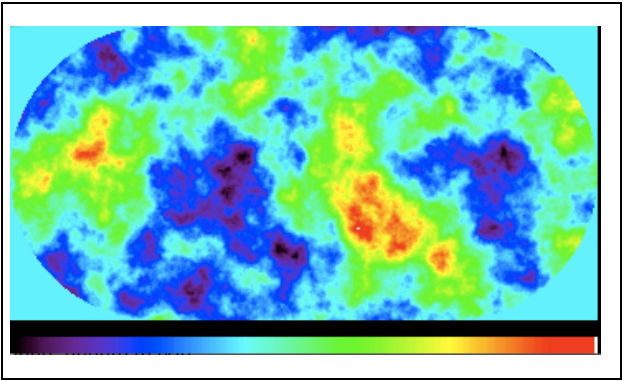

**Figure 13 NFIRAOS Turbulence phase screen**

For day time verification of NFIRAOS, it deploys a phase screen into the science optical path, shown as a green rectangular frame in Figure 2. Turbulence is also added blindly to DM commands. This concept eliminates a separate turbulence simulator including its light sources and telescope simulator in front of the entrance window. An important consideration is to reproduce median Mauna Kea $r_0$ & $\theta_2$.

We are investigating magneto-rheological polishing (MRF) of the phase screens, which are ~ 360 x 750 mm. The peak-to-valley, ~ 5 $\mu$m, of the wavefront OPD from the phase screen, and hence the material removal rate during polishing, and resulting cost of the phase screen depends critically on the measured $C_n^2$ profiles from Mauna Kea.

The optimal conjugation altitude and strength of the phase screen was determined via simulations based on site survey data.

## 13. SODIUM DENSITY AND ALTITUDE

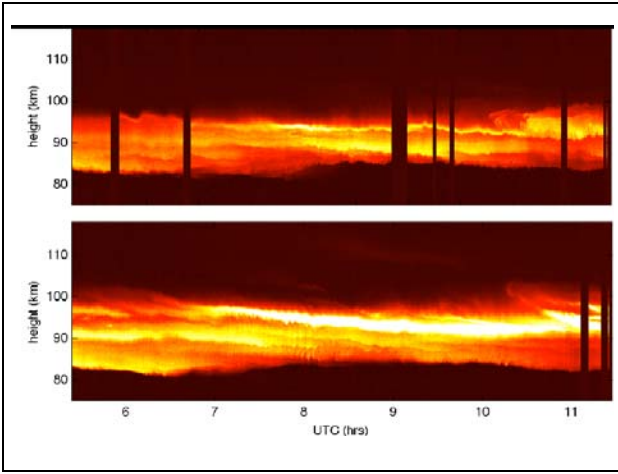 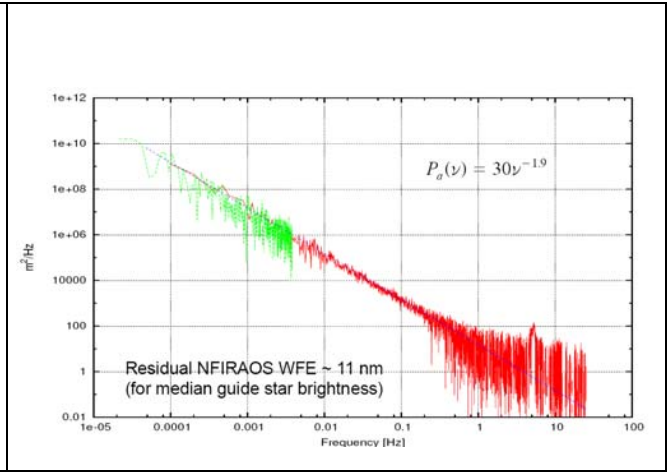

**Figure 14 Sodium Density profiles - UBC Vancouver**    **Figure 15 Power spectrum of Sodium Altitude**

Figure 14 Shows two different nights of data taken with a Lidar operated by the University of British Columbia using its 6-m Large Zenith Telescope operated just outside of Vancouver, Canada. Each column of pixels is a vertical profile through the mesospheric sodium layer. The horizontal axis covers about 6 hours of data. The dark vertical bands occurred when the Lidar was shut down due to the approach of an aircraft. Time varying structure resulting from turbulence and gravity waves is apparent. The bright horizontal bands that appear and drift downwards are from fresh sodium introduced by meteors.

These sodium variations perturb the wavefront sensor spot centroiding process. Since the outer subapertures of a Shack-Hartmann wavefront sensor have an oblique view of the laser as it traverses the sodium layer, the gain and zero points for each spot must be optimized by background processes, which take time. The resulting measurement errors and lags were evaluated via simulations, both in computer simulations, and on the adaptive optics test bench at the University of Victoria. Sodium movies were played into simulations, to assess: residual errors from meteor transients; power consumption of focusing trombone; determine a suitable update interval for background tasks; and residual errors from sodium variability

As well as the internal structural changes, the mean altitude of the sodium layer changes. On short time scales, this altitude change is interpreted as a focus error, but on longer time scales, it is corrected by refocusing the laser wavefront sensors by means of the trombone shown in the lower right of Figure 2, using the focus measurements from the OIWFS as a reference for true focus. A change of mean sodium altitude by 1 m causes 4 nm of apparent defocus. Sometimes the altitude changes by more than a kilometer in a few seconds which would be disastrous without robust control algorithms. Figure 15 shows a power spectrum of mean altitude of focus. The green data is old measurements by smaller lidars that were only able to make measurements every 70s. The UBC Lidar operates at 50 Hz and took the data shown in red. The best fit to more than a hundred nights of data is shown as a straight line. Inputting this power spectrum into models of NFIRAOS focus and trombone control indicates a residual of 11 nm, which fits in the NFIRAOS error budget.

## 14. TIME VARIABILITY OF $R_0$

Time series of $r_0$ are used in simulations of:

- Background control tasks such as WFS centroid gain estimations
- Image smearing during long exposures to assess astrometry accuracy

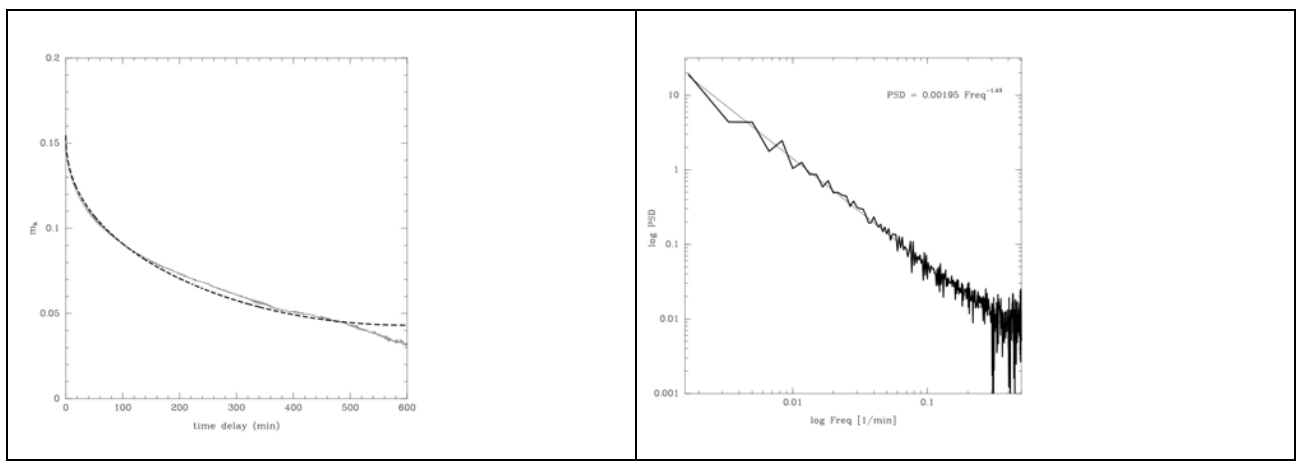

Figure 16 Autocorrelation of log($r_0$)        Figure 17 Power spectrum of $r_0$

However it is very difficult to choose a "representative" night time series. Periods of good seeing that last for hours are often followed by dramatic jumps in seeing on time scales of a few minutes. Searching through years of site testing data to find typical nights is nearly impossible. Therefore TMT has built an autoregressive model of seeing using data taken for more than a year. The autocorrelation of log($r_0$) is shown in Figure 16 for delays of up to 10 hours. The fourier transform gives the PSD of variations in log($r_0$) shown in the right figure. The best fit in log-log space is PSD = 1.95x10$^{-3}$ x f$^{1.43}$. Using this PSD, an autoregressive model with 100 taps was created. An example output time series generated by this model is shown in Figure 18.

Besides passing statistical tests, qualitatively, it displays many realistic features: Periods of good seeing that last for hours are often followed by dramatic jumps in seeing on time scales of a few minutes.

It would be desirable to extend this process to model the time evolution of layer strength in the $C_n^2$ profile. Such a model would be useful to assess the importance of good initial guess of layer strength for tomography, and how rapidly background tasks need to update estimated $C_n^2$ to provide priors for the real time tomography algorithm.

However, the technique for $r_0$ just described does not work for individual layers of TMT site data. The TMT data is too noisy per-layer. It even contains negative numbers sometimes. During the Kislovodsk conference, Marc Sarazin suggested that physical *priors* be applied as constraints in the MASS/DIMM data reduction. For example turbulence does not jump from layer to layer rapidly.

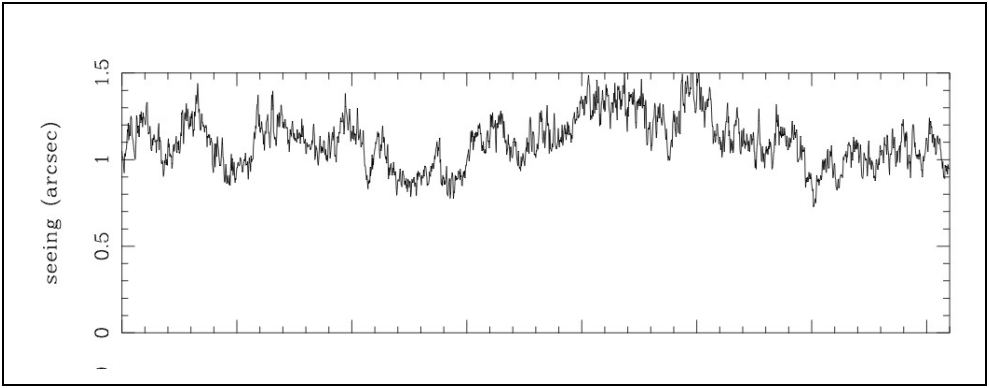

**Figure 18 AR model of bad seeing – 1 hour duration**

## 15. LASER LAUNCH TELESCOPE LOCATION AND DIAMETER

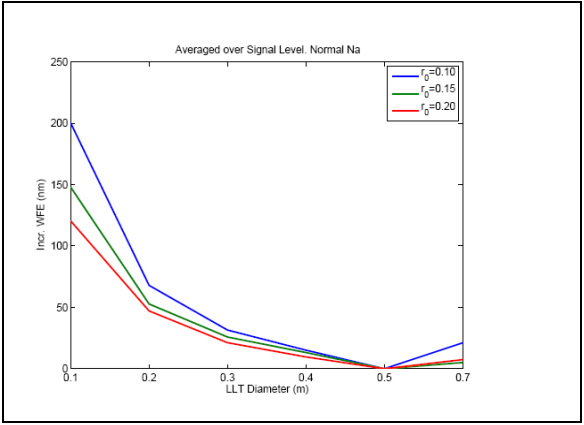

**Figure 19 Incremental wavefront error versus Laser launch telescope diameter**

End-to-end Monte Carlo physical optics simulations were conducted to determine whether to launch the lasers from the edge of the primary mirror as ELT plans, or from behind the secondary mirror. Side launch provides ~20 nm better Wavefront error, but at increased cost and complexity.

Figure 19 shows the change in wavefront error for various laser launch telescope diameters compared with the baseline 0.5 m diameter This simulation was run for the nominal sodium profile and Mauna Kea $C_n^2$ profile, for good, median and poor seeing of $r_0 = 20$, 15, and 10 cm.

## 16. SUMMARY

This paper has illustrated many ways in which site survey data is used to design and assess the performance of adaptive optics systems, as well as how site survey data is incorporated into the control system as *priors* to assist speed and accuracy of correcting atmospheric turbulence.

## ACKNOWLEDGEMENTS

The authors gratefully acknowledge the support of the TMT partner institutions. They are the Association of Canadian Universities for Research in Astronomy (ACURA), the California Institute of Technology and the University of California. This work was supported as well by the Gordon and Betty Moore Foundation, the Canada Foundation for Innovation, the Ontario Ministry of Research and Innovation, the National Research Council of Canada, the Natural Sciences and Engineering Research Council of Canada, the British Columbia Knowledge Development Fund, the Association of Universities for Research in Astronomy (AURA) and the U.S. National Science Foundation.

# REFERENCES

[1] G. Sanders and J. Nelson, "The status of the Thirty Meter Telescope project", Proc. SPIE 7733-69 (2010).
[2] B. Ellerbroek, "First light adaptive optics systems and components for the Thirty Meter Telescope," Proc. SPIE 7736-03 (2010).
[3] G. Herriot, et al., "NFIRAOS: facility adaptive optics system for the TMT", Proc. SPIE 7736-09 (2010).
[4] J.-P. Véran, C. Irvin, and G. Herriot, "Implementation of type-II tip-tilt control in NFIRAOS, with woofer-tweeter and vibration cancellation," Proc. SPIE 7736-167 (2010).
[5] L. Gilles, "Modeling update for the Thirty Meter Telescope laser guide star dual conjugate adaptive optics system", Proc. SPIE 7736-31 (2010).
[6] R. Conan, J.-P. Véran, and K. J. Jackson, "Experimental validation of type-II tip-tilt control in a woofer-tweeter adaptive optics system," Proc. SPIE 7736-169 (2010).
[7] D.-S. Moon, " The Science Calibration System for the TMT NFIRAOS and Client Instruments: Requirements and Design Studies," Proc. SPIE 7736, (2010).
[8] S. A. Wright, "The infrared imaging spectrograph (IRIS) for TMT: sensitivities and simulations", Proc.SPIE 7735-284 (2010).
[9] L. Wang, et al., "Impact of laser guide fratricide on TMT MCAO system", Proc. SPIE 7736-16 (2010).
[10] D. S. Hale, et al., "NIR low-order wavefront sensor for TMT IRIS", Proc. SPIE 7736-36 (2010).
[11] K. Vogiatzis, "Thermal modeling environment for TMT", Proc. SPIE 7738-11 (2010).
[12] K. Vogiatzis, "Unsteady aerodynamic simulations for TMT primary mirror segment wind loading", Proc. SPIE 7738-61 (2010).
[13] L. Poyneer, "Laboratory performance of the Gemini planet imager's adaptive optics wavefront reconstruction and control system", 7736-+86 (2010).